\documentclass[aps,pra,twocolumn,groupedaddress]{revtex4-1}

\usepackage{graphicx}
\usepackage{color}
\usepackage{bm}
\usepackage{amsmath}
\usepackage{amssymb}
\usepackage{braket}

\usepackage[utf8]{inputenc}

\newcommand{\eref}[1]{(\ref{#1})}

\begin{document}

\title{Theory of filtered type-II PDC in the continuous-variable domain: \\ Quantifying the impacts of filtering}

\author{Andreas Christ}
\email{andreas.christ@uni-paderborn.de}
\affiliation{Applied Physics and CeOPP, University of Paderborn, Warburger Stra{\ss}e 100, D-33098 Paderborn}
\author{Cosmo Lupo}
\affiliation{Research Laboratory of Electronics, Massachusetts Institute of Technology, Cambridge, Massachusetts, USA}
\author{Matthias Reichelt}
\author{Torsten Meier}
\affiliation{Department of Physics and CeOPP, University of Paderborn, Warburger Stra{\ss}e 100, D-33098 Paderborn, Germany}
\author{Christine Silberhorn}
\affiliation{Applied Physics and CeOPP, University of Paderborn, Warburger Straße 100, D-33098 Paderborn}


\begin{abstract}
Parametric down-conversion (PDC) forms one of the basic building blocks for quantum optical experiments. However, the intrinsic multimode spectral-temporal structure of pulsed PDC often poses a severe hindrance for the direct implementation of the heralding of pure single-photon states or, for example, continuous-variable entanglement distillation experiments. To get rid of multimode effects narrowband frequency filtering is frequently applied to achieve a single-mode behavior.

A rigorous theoretical description to accurately describe the effects of filtering on PDC, however, is still missing. To date, the theoretical models of filtered PDC are rooted in the discrete-variable domain and only account for filtering in the low gain regime, where only a few photon pairs are emitted at any single point in time. In this paper we extend these theoretical descriptions and put forward a simple model, which is able to accurately describe the effects of filtering on PDC in the continuous-variable domain.

This developed straightforward theoretical framework enables us to accurately quantify the trade-off between suppression of higher-order modes, reduced purity and lowered Einstein-Podolsky-Rosen (EPR) entanglement, when narrowband filters are applied to multimode type-II PDC.
\end{abstract}


\maketitle

\section{Introduction}
Since the landmark experiment by Hong-Ou-Mandel in 1987 \cite{hong_measurement_1987} parametric down-conversion (PDC) has become one of the most widely used basic building blocks for quantum optical experiments. For example, it serves as a source of entangled photon pairs \cite{kwiat_new_1995, kwiat_ultrabright_1999, kurtsiefer_high-efficiency_2001, herrmann_post-selection_2013, pan_multiphoton_2012}, enables the heralding of single photon states \cite{mosley_heralded_2008, uren_efficient_2004, pittman_heralding_2005, migdall_tailoring_2002, krapick_efficient_2013} and the generation of Einstein-Podolsky-Rosen (EPR) entanglement \cite{eckstein_highly_2011, kurochkin_distillation_2014}.

However, PDC suffers from one major drawback. Standard PDC sources do not emit their quantum states in a single, well defined, optical mode, but into a multitude of different spectral-temporal and spatial modes simultaneously. While this can be beneficial for multiplexing purposes \cite{christ_exponentially_2012} or multimode detection schemes \cite{armstrong_programmable_2012}, to date, most experiments require single-mode quantum states. To get rid of multiple spatial modes in the generated PDC state waveguides can be used to achieve an emission into a single well-defined spatial mode \cite{mosley_direct_2009, christ_spatial_2009}.

The spectral multimode structure of PDC has been studied extensively and there exist two main approaches to achieve a spectral single-mode emission from PDC \textit{source engineering} and \textit{filtering}: \textit{source engineering} refers to a pulsed PDC process where the parameters are specifically adjusted to ensure an emission of the photon pairs directly into a single optical mode \cite{grice_spectral_1997, uren_generation_2006, mosley_heralded_2008, gerrits_generation_2011, eckstein_highly_2011, harder_optimized_2013, jin_nonclassical_2013}. However, this approach only works at specific wavelength ranges within certain materials. \textit{Filtering}, on the other hand, relies on placing narrowband spectral filters after the PDC process to effectively suppress the multimode structure. It is easy, robust and straightforward to implement, at the cost of introducing losses. To date filtered PDC sources have been used extensively for the heralding of pure single-photons and the generation of entangled photon pairs \cite{branczyk_optimized_2010, huang_heralding_2010, huang_optimized_2011, patel_erasing_2012, riedmatten_quantum_2003, kaltenbaek_experimental_2006}.

The effects of filters for photon-subtraction experiments have already been investigated \cite{tualle-brouri_multimode_2009}. However, a rigorous theoretical description of filtered PDC is still missing. While several theoretical models, rooted in the discrete-variable domain exist \cite{branczyk_optimized_2010, huang_heralding_2010, huang_optimized_2011, patel_erasing_2012}, these are concerned with the low gain regime. They are sufficient for the heralding of single-photons and entangled photon-pair generation, but are not adequate for quantum optics experiments in the continuous-variable domain such as entanglement distillation protocols \cite{opatrny_improvement_2000, takahashi_entanglement_2010, kurochkin_distillation_2014}.

Concerning PDC in the high gain regime or continuous variable domain, i.e., regarding PDC as a source of squeezed states, several theoretical descriptions already exist \cite{christ_probing_2011, christ_theory_2013-1, wasilewski_pulsed_2006, lvovsky_decomposing_2007, brambilla_simultaneous_2004, caspani_tailoring_2010, gatti_multiphoton_2003, dayan_theory_2007,bennink_improved_2002}. These, however, do not consider the effects of filters on the generated squeezing amplitudes, the multimode character, and the purity of the filtered state.

In principle the impacts of filtering on PDC are quite intuitive. By reshaping the spectrum of the generated photons the filter operation suppresses higher-order modes at the cost of reduced squeezing values and a lowered purity. In this paper we perform a formal analysis of this intuition and as a first step revisit the mathematical framework of PDC in the continuous-variable domain and the theoretical description of filter operations. By using these we develop a simple and straightforward theoretical model of filtered PDC in the continuous-variable domain, which enables us to accurately quantify impacts of filtering. Finally, we use this model to analyze the trade-off between the suppression of higher-order modes, the remaining EPR squeezing, and purity. To simplify the discussion we restrict ourselves to type-II PDC processes.

\section{Overview}
\begin{figure}[htb]
    \begin{center}
        \includegraphics[width=0.75\linewidth]{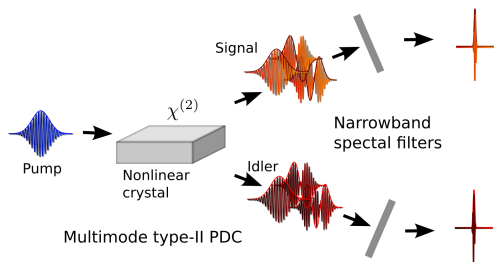}
    \end{center}
    \caption{Narrowband filtering of a broadband multimode type-II PDC process, followed by a mode optimization, enables the creation of EPR-entanglement in a single optical mode.}
    \label{fig:erq_filtering_overview}
\end{figure}

Fig. \ref{fig:erq_filtering_overview} sketches the overall filtering process: We start with a standard type-II PDC source, which emits a multitude of finitely squeezed EPR states in broadband spectral modes. We then apply narrowband spectral filtering in both the signal and the idler arm to get rid of the intrinsic multimode spectral structure. 

This paper is structured into four main parts: In Sec. \ref{sec:type-II_PDC} we review the process of type-II PDC and discuss the properties of the generated quantum states. In Sec. \ref{sec:filtering} we mathematically describe the filtering process and discuss its impact on type-II PDC. The main result of our research is presented in Sec. \ref{sec:epr-entanglement_generation}. In this section we show that a basis transformation from the original broadband spectral modes (\textit{Schmidt modes}) to a new mode set adapted to the applied filtering (\textit{effective Schmidt modes}) enables us to accurately gauge the effects of the filtering on the PDC state; to be precise the suppression of higher-order modes, the remaining EPR squeezing, and the lowered purity. To perform this basis optimization we introduce two approaches: a simple and straightforward model useful for designing filtered PDC sources, and a complicated rigorous model  which verifies that our straightforward approach is indeed optimal. Finally, in Sec. \ref{sec:analysis} we use our developed framework to accurately quantify the trade-off between higher-order mode suppression, remaining EPR squeezing, and the purity of the resulting state.

\section{Type-II PDC}\label{sec:type-II_PDC}
During the process of type-II PDC a photon of an incoming pump beam spontaneously decays, inside a crystal featuring a \(\chi^{(2)}\) nonlinearity, into a photon-pair usually labeled signal and idler, where signal and idler exhibit orthogonal polarizations. In the scope of this paper we consider the strong pumping regime, where several photon pairs are created simultaneously, and the emitted photon pairs form EPR states.

Mathematically the generated type-II PDC state can be described as \cite{christ_probing_2011,christ_theory_2013-1}
\begin{align}
    \nonumber
    \ket{\psi}_{PDC} &=  
    \exp\left[-\frac{\imath}{\hbar}\left(B \iint \mathrm d \omega_s \, \mathrm d \omega_i \, f(\omega_s, \omega_i) \hat{a}^\dagger(\omega_s) \hat{b}^\dagger(\omega_i) \right. \right. \\
    &\left. \left. + H.c. \right) \right] \ket{0}.
    \label{eq:pdc_state}
\end{align}
Here \(\hat{a}^\dagger(\omega_s)\), the photon creation operator, describes the generation of a signal photon at frequency \(\omega_s\) and \(\hat{b}^\dagger(\omega_i)\) describes the generation of an idler photon at frequency \(\omega_i\). The function \(f(\omega_s, \omega_i)\) is the joint-spectral amplitude (JSA) of the emitted photon pairs, which in general exhibits correlations between \(\omega_s\) and \(\omega_i\). \(B\) collects all constants and is often referred to as the optical gain which describes the efficiency of the process.

Eq. \eref{eq:pdc_state} does not directly reveal the quantum properties of the emitted type-II PDC-state. In order to obtain these we have to perform a singular-value decomposition or \textit{Schmidt decomposition} \cite{law_continuous_2000, schmidt_zur_1907, bunse-gerstner_singular_1988, nielsen_quantum_2004} of the JSA, which allows us to express the state in terms of pulsed pairs of uncorrelated states with specific broadband spectra. To be precise, we decompose the exponent in Eq. \eref{eq:pdc_state} into a sum of positive amplitudes \(r_k\) and broadband mode functions \(\psi_k(\omega_s)\) and \(\phi_k(\omega_i)\) \cite{christ_probing_2011, christ_theory_2013-1}
\begin{align}
    \nonumber
    -\frac{\imath}{\hbar} B f(\omega_s, \omega_i) &= \sum_k r_k \psi^*_k(\omega_s) \phi^*_k(\omega_i)\\
    -\frac{\imath}{\hbar} B^* f^*(\omega_s, \omega_i) &= - \sum_k r_k \psi_k(\omega_s) \phi_k(\omega_i).
    \label{eq:schmidt_decomposition}
\end{align}

\begin{figure}[htb]
    \begin{center}
        \includegraphics[width=0.95\linewidth]{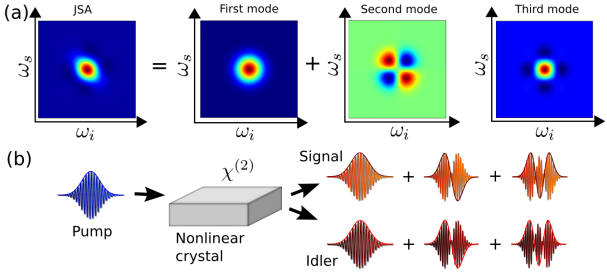}
    \end{center}
    \caption{Visualization of a \textit{Schmidt decomposition}: Each pair of \textit{Schmidt modes} of signal and idler, depicted in panel (b), forms a spectral distribution, which, combined and weighted by their \(r_k\) values, forms the JSA in panel (a).}
    \label{fig:schmidt_decomposition}
\end{figure}

Each pair of \textit{Schmidt modes} \(\psi_k(\omega_s)\) and \(\phi_k(\omega_s)\) defines a spectral distribution which, weighted by their individual amplitudes \(r_k\) yields the JSA. We visualized this connection in Fig. \ref{fig:schmidt_decomposition} for a PDC state consisting of three optical modes with equal weights \(r_k\). The individual elements of the sum, which create the JSA, are depicted in Fig. \ref{fig:schmidt_decomposition}(a) and the respective mode functions are depicted in Fig. \ref{fig:schmidt_decomposition}(b). In general type-II PDC experiments, the JSA is very close to a two-dimensional Gaussian and the resulting \textit{Schmidt modes} are extremely similar to the Hermite functions \cite{uren_photon_2003}.

By using these \textit{Schmidt modes} we are able to define the broadband single-photon operators \cite{rohde_spectral_2007}
\begin{align}
    \nonumber
    \hat{A}_k = \int \mathrm d \omega_s \, \psi_k(\omega_s) \hat{a}(\omega_s)\\
    \hat{B}_k = \int \mathrm d \omega_i \, \phi_k(\omega_i) \hat{b}(\omega_i),
    \label{eq:broadband_modes}
\end{align}
i.e. we introduce the single-photon creation operators \(\hat{A}_k^\dagger\) and \(\hat{B}_k^\dagger\), which create photons not at a single-frequency \(\omega\), but in a broadband frequency range. By using Eqs. \eref{eq:schmidt_decomposition} and \eref{eq:broadband_modes} we are able to rewrite the PDC state from Eq. \eref{eq:pdc_state} as \cite{christ_probing_2011}
\begin{align}
    \nonumber
    \ket{\psi}_{PDC} &= \exp\left[\sum_k r_k \hat{A}_k^\dagger \hat{B}_k^\dagger - H.c.\right]\ket{0}\\
    \nonumber
    &= \bigotimes_k \exp\left[r_k \hat{A}_k^\dagger \hat{B}_k^\dagger - H.c.\right]\ket{0}\\
    &= \bigotimes_k \hat{S}_{AB}(-r_k)\ket{0}.
    \label{eq:pdc_state_broadband}
\end{align}
From Eq. \eref{eq:pdc_state_broadband} it is evident that type-II PDC emits a multitude of finitely squeezed EPR states \( \bigotimes_k \hat{S}_{AB}(-r_k)\ket{0}\) \cite{barnett_methods_2003}. The EPR squeezing is, however, not created at single frequencies but in the broadband modes \(\hat{A}_k\) and \(\hat{B}_k\). From a physical point of view this means that type-II PDC creates a multitude of optical pulses which contain finitely squeezed EPR states, as already schematically depicted in Fig. \ref{fig:erq_filtering_overview}.

This property of type-II PDC is only visible when using the obtained \textit{Schmidt modes}, and the individual optical modes are, solely in this basis, completely independent from each other. These broadband modes are also referred to as the \textit{Schmidt basis} or \textit{Eigenbasis} of the system. In terms of dB the EPR squeezing in mode \(k\) is defined as \cite{christ_probing_2011}
\begin{equation}
    \mathrm{squeezing[dB]} = -10 \log_{10} \left(e^{-2 r_k}\right).
    \label{eq:squeezing_definition}
\end{equation}
In the Heisenberg picture we are able to write the type-II PDC process as \cite{christ_theory_2013-1}
\begin{align}
    \nonumber
    \hat{a}^{(\mathrm{out})}(\omega) &= \int \mathrm d \omega' \, U_a(\omega, \omega') \hat{a}^{(\mathrm{in})}(\omega') + V_a(\omega, \omega') \hat{b}^{(\mathrm{in})\dagger}(\omega') \\
    \hat{b}^{(\mathrm{out})}(\omega) &= \int \mathrm d \omega' \,  U_b(\omega, \omega') \hat{b}^{(\mathrm{in})}(\omega') + V_b(\omega, \omega') \hat{a}^{(\mathrm{in})\dagger}(\omega').
    \label{eq:pdc_state_heisenberg}
\end{align}
Here we added the labels \( \mathrm{(in)} \) and \( \mathrm{(out)} \) to the single-photon creation and destruction operators to stress that the modes on the left, labeled \( \mathrm{(out)} \), are the modes after the PDC process has taken place. They are given as a function of the modes \( \mathrm{(in)} \) which label the modes before the PDC process. The \(U_{a,b}(\omega, \omega')\) and \(V_{a,b}(\omega, \omega')\) matrices contain the process properties. They are of the form \cite{christ_theory_2013-1}
\begin{align}
    \nonumber
    U_a(\omega, \omega') = \sum_k \psi_k^*(\omega) \cosh(r_k) \psi_k(\omega') \\
    \nonumber
    V_a(\omega, \omega') = \sum_k \psi_k^*(\omega) \sinh(r_k) \phi_k^*(\omega') \\
    \nonumber
    U_b(\omega, \omega') = \sum_k \phi_k^*(\omega) \cosh(r_k) \phi_k(\omega')\\
    V_b(\omega, \omega') = \sum_k \phi_k^*(\omega) \sinh(r_k) \psi_k^*(\omega').
    \label{eq:U_V_matrices}
\end{align}
We are able to cast Eq. \eref{eq:pdc_state_heisenberg} into the broadband mode formalism using the broadband single-photon creation and destruction operators defined in Eq. \eref{eq:broadband_modes}. We arrive at
\begin{align}
    \nonumber
    \hat{A}_k^{(\mathrm{out})} &= \cosh(r_k) \hat{A}_k^{(\mathrm{in})} + \sinh(r_k) \hat{B}_k^{(\mathrm{in})\dagger} \\
    \hat{B}_k^{(\mathrm{out})} &= \cosh(r_k) \hat{B}_k^{(\mathrm{in})} + \sinh(r_k) \hat{A}_k^{(\mathrm{in})\dagger}.
    \label{eq:pdc_state_heisenberg_broadband}
\end{align}
Again using the \textit{Schmidt modes} directly reveals the EPR-squeezing properties of the process. From a physical point of view the output modes \(\hat{A}_k^{(\mathrm{out})}\) and \(\hat{B}_k^{(\mathrm{out})}\), are the modes in which we observe or measure our quantum state. For example using the modes \(\hat{A}_k^{(\mathrm{out})}\) and \(\hat{B}_k^{(\mathrm{out})}\) for homodyne detection would reveal the EPR entanglement, whereas a measurement in a different basis hides the entanglement.

\begin{figure}[htb]
    \begin{center}
        \includegraphics[width=0.8\linewidth]{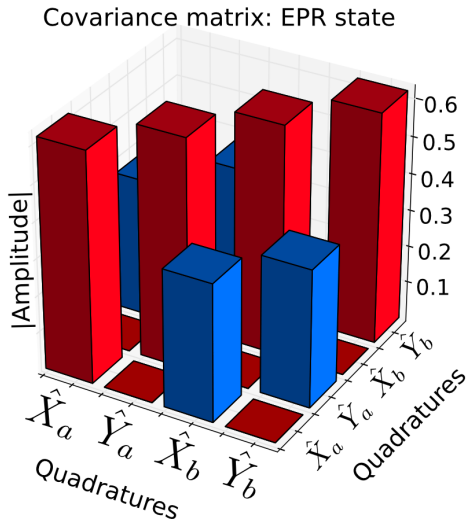}
    \end{center}
    \caption{Covariance matrix of an EPR state with 3\(\,\mathrm{dB}\) of squeezing. Plotted is the absolute value of the individual elements. The \(\sinh(r)\) side peaks are characteristic for EPR-entanglement.}
    \label{fig:epr_cov_mat}
\end{figure}

For our purposes it is very useful to consider the covariance matrix of the generated type-II PDC states by using the obtained \textit{Schmidt modes} \cite{christ_exponentially_2012}. In order to calculate the individual covariance matrix elements we first define the broadband quadrature operators for the individual optical modes:
\begin{align}
    \nonumber
    &\hat{X}_a^k = \frac{1}{\sqrt{2}}\left(\hat{A}_k + \hat{A}_k^\dagger\right) \qquad
    \hat{X}_b^k = \frac{1}{\sqrt{2}}\left(\hat{B}_k + \hat{B}_k^\dagger\right)\\
    &\hat{Y}_a^k = \frac{1}{\sqrt{2}\imath}\left(\hat{A}_k - \hat{A}_k^\dagger\right) \qquad
    \hat{Y}_b^k = \frac{1}{\sqrt{2}\imath}\left(\hat{B}_k - \hat{B}_k^\dagger\right).
    \label{eq:broadband_quadratures}
\end{align}
Here the label \(k\) depicts the number of the individual optical mode. For one EPR state we required \(\hat{X}_a^k\) and \(\hat{Y}_a^k\) to describe the signal mode and \(\hat{X}_b^k\) and \(\hat{Y}_b^k\) for the idler mode. For a PDC state consisting of \(N\) EPR states we can group the \( 4 \times N\) quadrature operators in the vector
\begin{align}
    \vec{R} = \left(\hat{X}_a^1, \hat{Y}_a^1, \hat{X}_b^1, \hat{Y}_b^1, \dots, \hat{X}_a^N, \hat{Y}_a^N, \hat{X}_b^N, \hat{Y}_b^N\right),
    \label{eq:quadrature_vector}
\end{align}
i.e., the first four elements of the vector consist of the four quadrature operators for the first optical mode defined by \(\hat{A}_1\) and \(\hat{B}_1\), followed by the four quadrature operators for the second mode, and so forth until mode \(N\). In terms of \(\vec{R}\) the individual covariance elements \(\sigma_{ij}\) are defined as \cite{adesso_entanglement_2007}
\begin{align}
    \sigma_{ij} = \frac{1}{2} \langle \hat{R}_i \hat{R}_j + \hat{R}_j \hat{R}_i \rangle - \langle \hat{R}_i \rangle \langle \hat{R}_j \rangle.
    \label{eq:covariance_elements}
\end{align}
The covariance matrix \(\sigma\) for a single EPR state with an EPR squeezing amplitude \(r\) is given by \cite{adesso_entanglement_2007}
\begin{align}
    \sigma = \frac{1}{2}
    \begin{pmatrix}
        \cosh(2r) & 0 & \sinh(2r) & 0 \\
        0 & \cosh(2r) & 0 & -\sinh(2r) \\
        \sinh(2r) & 0 & \cosh(2r) & 0 \\
        0 & -\sinh(2r) & 0 & \cosh(2r)
    \end{pmatrix}.
    \label{eq:cov_mat_pdc}
\end{align}
We visualized the covariance matrix of an EPR state containing 3\(\, \mathrm{dB}\) of EPR squeezing in Fig. \ref{fig:epr_cov_mat}, where we plotted the absolute value of the individual elements. Characteristic for EPR-entanglement are the side peaks from the \(\sinh(r)\) terms.

\begin{figure}[htb]
    \begin{center}
        \includegraphics[width=0.90\linewidth]{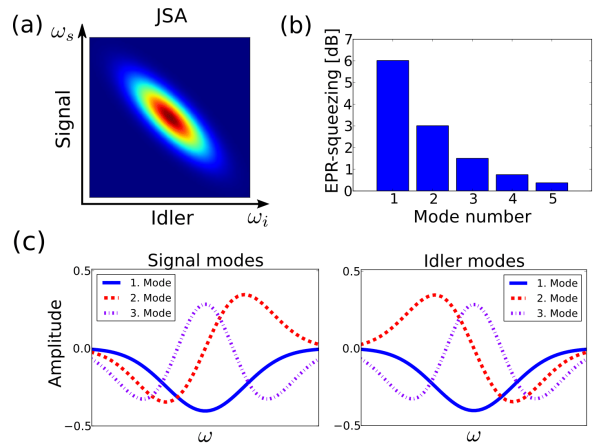}
    \end{center}
    \caption{PDC state used to demonstrate the individual steps of our protocol: (a) the  JSA \(f(\omega_s, \omega_i)\), (b) First five EPR-squeezing values, (c) First three optical modes of the signal and idler beams.}
    \label{fig:pdc_example}
\end{figure}

In the following we are going to demonstrate the impact of filtering on the exemplary pulsed PDC state depicted in Fig. \ref{fig:pdc_example}, which exhibits correlations in frequency and thus several broadband modes. Here, Fig. \ref{fig:pdc_example} (a) shows its respective JSA \(f(\omega_s, \omega_i)\) and Fig. \ref{fig:pdc_example} (b) shows the EPR squeezing in the first five modes. We adjusted the optical gain such that the first mode has an EPR squeezing of about 6\(\,\mathrm{dB}\). Fig. \ref{fig:pdc_example} (c) depicts the first three signal and idler modes. Note, that we explicitly choose a real-valued JSA distribution leading to real valued \textit{Schmidt modes}, which facilitate a straightforward visualization throughout the paper. The exact simulation parameters are given in App. \ref{app:simulated_pdc_state}.

\begin{figure}[htb]
    \begin{center}
        \includegraphics[width=0.9\linewidth]{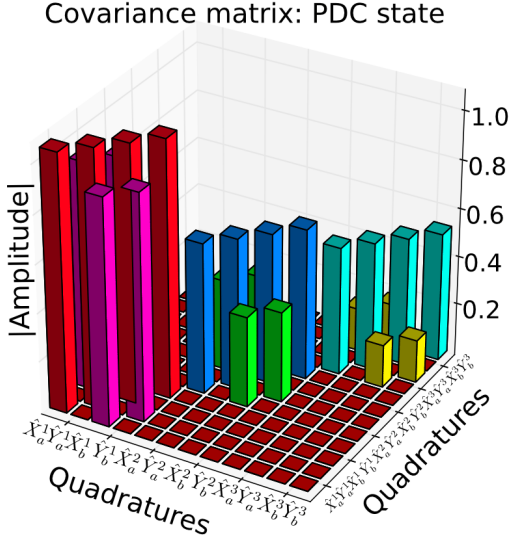}
    \end{center}
    \caption{Covariance matrix of the unfiltered PDC state depicted in Fig. \ref{fig:pdc_example}. This figures shows the absolute values of the individual elements for the first three optical modes. The submatrix for each mode is identical to a finitely squeezed EPR state.}
    \label{fig:cov_mat_unfilt}
\end{figure}

We plotted the corresponding covariance matrix of our exemplary type-II PDC state in Fig. \ref{fig:cov_mat_unfilt}. Here we restricted ourselves to a visualization of the first three optical modes and again plotted the absolute values of the individual elements. Each submatrix is identical to a finitely squeezed EPR state. Note that all \(\sigma_{ij}\) elements between different optical modes are zero, indicating that the individual modes are completely independent from each other.

\section{Filtering}\label{sec:filtering}
In the context of photon-pair generation and the heralding of pure single-photons, the effects of filters on type-II PDC have already been studied extensively, \cite{branczyk_optimized_2010, huang_heralding_2010, huang_optimized_2011, patel_erasing_2012, riedmatten_quantum_2003, kaltenbaek_experimental_2006}. In the scope of this paper we extend this analysis to the continuous-variable domain and the corresponding EPR-state generation via type-II PDC.

The filtering has two main effects on the PDC state: First, it introduces losses, i.e., we couple vacuum into the system and degrade its purity and the EPR entanglement. Second, the filtering reshapes the frequency spectra, destroying the original mode structure.

To mathematically describe this process, we model a spectral filter acting upon an optical field as a frequency-dependent loss channel or beam splitter \cite{branczyk_optimized_2010}:
\begin{align}
    \hat{a}^{(\mathrm{out})}(\omega) = T(\omega)\hat{a}^{(\mathrm{in})}(\omega) + R(\omega)\hat{v}(\omega), 
    \label{eq:filter_operation}
\end{align}
where \(\left|T(\omega)\right|^2\) is the transmission and \(\left|R(\omega)\right|^2\) the reflection probability at frequency \(\omega\). They obey \(\left|T(\omega)\right|^2 + \left|R(\omega)\right|^2 = 1\). The operator \(\hat{v}(\omega)\) describes the vacuum introduced into the system.

In the Schrödinger picture --- or photon-number representation --- it is, in general, very difficult to describe filtered PDC states, since we have to consider a frequency-dependent loss and further have to take into account all possible combinations of photons being reflected and transmitted at the filter. In our description we work in the continuous-variable domain and simply apply Eq. \eref{eq:filter_operation} on Eq. \eref{eq:pdc_state_heisenberg} to obtain the filtered PDC state in the Heisenberg picture. In this way we avoid having to evaluate the exact photon-number properties and arrive at
\begin{align}
    \nonumber
    \hat{a}^{(\mathrm{out})}(\omega) &= T_a(\omega)\left[\int \mathrm d \omega' \, U_a(\omega, \omega') \hat{a}^{(\mathrm{in})}(\omega') \right. \\
    \nonumber
    &\left. + V_a(\omega, \omega')\hat{b}^{(\mathrm{in})\dagger}(\omega')\right] + R_a(\omega) \hat{v}_a(\omega) \\
    \nonumber
    \hat{b}^{(\mathrm{out})}(\omega) &= T_b(\omega)\left[\int \mathrm d \omega' \, U_b(\omega, \omega') \hat{b}^{(\mathrm{in})}(\omega') \right. \\
    &\left.+ V_b(\omega, \omega')\hat{a}^{(\mathrm{in})\dagger}(\omega')\right] + R_b(\omega) \hat{v}_b(\omega).
    \label{eq:pdc_state_filtered}
\end{align}
Here, \(T_a(\omega)\) and \(R_a(\omega)\) describe the action of a filter in the signal arm, \(T_b(\omega)\) and \(R_b(\omega)\) the action of the filter in the idler arm, and \(\hat{v}_a\) and \(\hat{v}_b\) label the added vacuum contributions in the signal and idler arms, respectively.

Similar to the previous section we cast Eq. \eref{eq:pdc_state_filtered} in the broadband mode picture. However, this is more difficult than in the previous section, since we have to consider the effects of frequency dependent loss on the PDC state. 

From a physical point of view we can think of this filtering as a reshaping of the individual optical modes, where effectively each optical mode is multiplied by the filter function. The corresponding destruction of the orthogonality between different modes leads to intermodal couplings, and cross correlations appear. From a mathematical point of view, filtered PDC states are a special case of the well-known fact that mixed states, in general, do not have a Schmidt decomposition \cite{terhal_schmidt_2000, eisert_schmidt_2001, aniello_class_2008}. It is, in general, not possible any more to find a \textit{Schmidt basis} or \textit{eigenbasis}; this means an orthogonal basis for the system, where all optical modes are independent from each other. A detailed mathematical discussion of this effect is given in App. \ref{app:filtered_type-II_PDC}, where we prove that it is, for all practical purposes, impossible to find a new \textit{Schmidt basis} for the system.

We consequently introduce the two new broadband-mode sets, to be able to express the state into a new arbitrary basis, which we label
\begin{align}
\hat{C}_k &= \int \mathrm d \omega \, f_k(\omega) \hat{a}(\omega) \label{eq:new_broadband_c}, \\
\hat{D}_k &= \int \mathrm d \omega \, g_k(\omega) \hat{b}(\omega).                  \label{eq:new_broadband_d}
\end{align}
To be precise, the new mode sets \(\{f_k(\omega)\}\) and \(\{g_k(\omega)\}\) each have to form a complete and orthogonal basis set.
By using these new basis sets we can express the filtered type-II PDC state as
\begin{align}
    \nonumber
    \hat{C}_k^{(\mathrm{out})} &= \int \mathrm d \omega' \, U_a^k(\omega') \hat{a}^{(\mathrm{in})}(\omega') + \int \mathrm d \omega' \, V_a^k(\omega') \hat{b}^{(\mathrm{in})\dagger}(\omega') \\
    \nonumber
    &+ \int \mathrm d \omega \, R_a^k(\omega) \hat{v}_a(\omega) \\
    \nonumber
    \hat{D}_k^{(\mathrm{out})} &= \int \mathrm d \omega' \, U_b^k(\omega') \hat{b}^{(\mathrm{in})}(\omega') + \int \mathrm d \omega' \, V_b^k(\omega') \hat{a}^{(\mathrm{in})\dagger}(\omega') \\
    &+ \int \mathrm d \omega \, R_b^k(\omega) \hat{v}_b(\omega),
    \label{eq:pdc_state_filtered_broadband}
\end{align}
with \(U_a^k(\omega'), U_b^k(\omega'), V_a^k(\omega'), V_b^k(\omega'), R_a^k(\omega) \, \mathrm{and} \, R_b^k(\omega)\) defined as
\begin{align}
    \nonumber
    U_a^k(\omega') &= \int \mathrm d \omega \, f_k(\omega) T_a(\omega) U_a(\omega, \omega') \\
    \nonumber
    U_b^k(\omega') &= \int \mathrm d \omega \, g_k(\omega) T_b(\omega) U_b(\omega, \omega') \\
    \nonumber
    V_a^k(\omega') &= \int \mathrm d \omega \, f_k(\omega) T_a(\omega) V_a(\omega, \omega') \\
    \nonumber
    V_b^k(\omega') &= \int \mathrm d \omega \, g_k(\omega) T_b(\omega) V_b(\omega, \omega') \\
    R_a^k(\omega) &= f_k(\omega) R_a(\omega) \qquad
    R_b^k(\omega) = g_k(\omega) R_b(\omega).
    \label{eq:pdc_state_filtered_broadband_definitions}
\end{align}
These formulas enable us to study the properties of the filtered type-II PDC states into a variety of different broadband mode sets. From a physical point of view the new basis sets \(\hat{C}_k^{\mathrm{(out)}}\) and  \(\hat{D}_k^{\mathrm{(out)}}\) define the modes in which we \textit{measure} the state. From a mathematical point of view we simply performed a basis transformation, with the aim to find a new basis adapted to the filtered state.

\begin{figure}[htb]
    \begin{center}
        \includegraphics[width=0.99\linewidth]{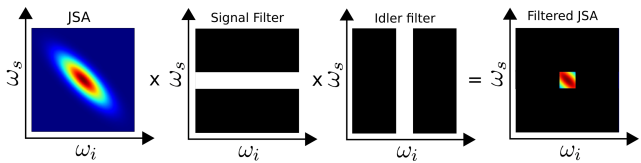}
    \end{center}
    \caption{A rectangular filter in the signal and idler arm cuts the JSA and lets only the central frequencies pass.}
    \label{fig:jsa_filt}
\end{figure}

In the scope of this paper we restrict ourselves to identical rectangular filter functions in the signal and idler, which enables us to precisely cut different parts of the spectrum from the PDC state. (Gaussian filters were tested as well and yielded similar results.) An exemplary rectangular filter in the signal and idler arm with respect to the JSA, introduced in Fig. \ref{fig:pdc_example}, is shown in Fig. \ref{fig:jsa_filt}. The filter cuts the JSA and only the central frequencies of the PDC photons are able to pass undisturbed.

\begin{figure}[htb]
    \begin{center}
        \includegraphics[width=0.99\linewidth]{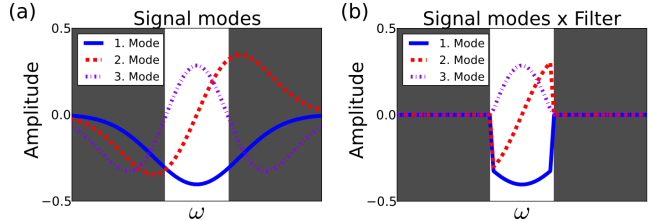}
    \end{center}
    \caption{(a) Signal modes with respect to the applied rectangular filter in Fig. \ref{fig:jsa_filt}. (b) Individual signal mode functions multiplied by the filter.}
    \label{fig:modes_filt}
\end{figure}

The filter with respect to the signal modes of the original PDC state is shown in Fig. \ref{fig:modes_filt}(a). Fig. \ref{fig:modes_filt} (b) presents the individual signal modes multiplied by the filter function.

These two figures already show that the impact of filtering on a PDC state is very different from standard losses. When an EPR state undergoes normal losses vacuum is added to the system and EPR squeezing is lost, but the mode structure remains unchanged. In the case of filtering vacuum is added as well, but the spectrally dependent losses also significantly alter the spectral structure of the PDC state. 

\begin{figure}[htb]
    \begin{center}
        \includegraphics[width=0.78\linewidth]{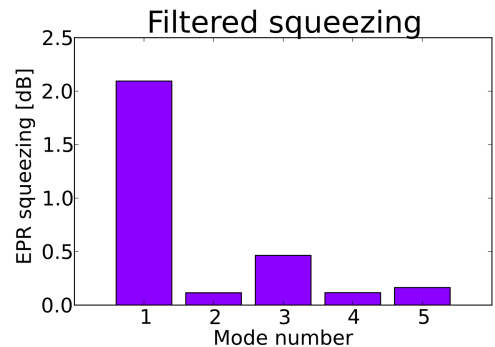}
    \end{center}
    \caption{The EPR squeezing in the first five modes  after the filtering has been applied. The EPR squeezing in all modes is significantly reduced.}
    \label{fig:filtered_epr_squeezing}
\end{figure}

We visualized the effect of the filtering on the EPR-squeezing amplitudes, using the original \textit{Schmidt basis}, in Fig. \ref{fig:filtered_epr_squeezing}. In comparison to the unfiltered EPR squeezing shown in Fig. \ref{fig:pdc_example} (b) all amplitudes are significantly reduced, which represents the losses introduced by the filters. (The formulas to calculate the filtered EPR squeezing are given in App. \ref{app:squeezing_after_filtering}.)

\begin{figure}[htb]
    \begin{center}
        \includegraphics[width=0.88\linewidth]{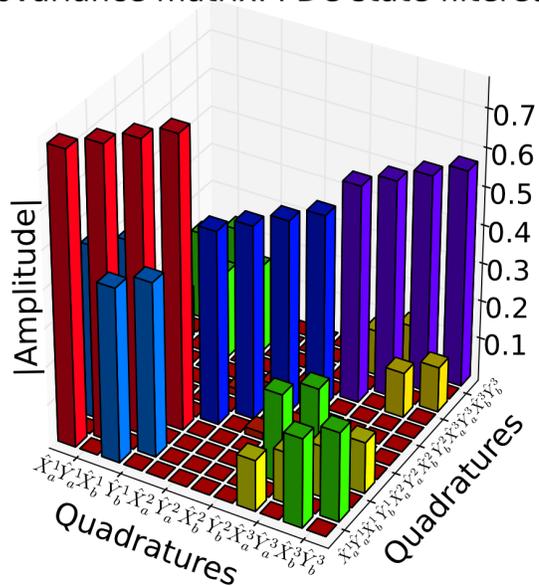}
    \end{center}
    \caption{Filtering a PDC state introduces losses, which move the state towards vacuum and further leads to correlations between the different optical modes.}
    \label{fig:pdc_cov_mat_filt}
\end{figure}

The impact of the filtering on the mode structure and the vacuum added to the system are also directly visible in the covariance matrix representation depicted in Fig. \ref{fig:pdc_cov_mat_filt}. Again we plotted the absolute value of the individual elements and use the broadband quadratures from Eq. \eref{eq:broadband_quadratures}. The exact formula for the full covariance matrix of a filtered PDC state is given in App. \ref{app:covariance_matrix}.

In comparison with the unfiltered PDC state depicted in Fig. \ref{fig:cov_mat_unfilt}, the individual amplitudes are significantly decreased, with the central peaks moving toward the vacuum amplitude of 0.5, i.e., the losses introduced by the filtering shifts the state towards vacuum. In contrast to standard losses, however, additional cross correlations between different optical modes appear, which are clearly visible between the first and third mode. These are a direct result of the reshaping of the spectral properties by the filter function.

Note that, in this specific scenario, there are no correlations with respect to the second mode, because the first and third mode are symmetric and the second mode is antisymmetric with respect to the origin. This property is not affected by the applied filtering, and consequently no couplings between these modes occur. This is, however, only true for this specific scenario. Filters which are not centered perfectly, will introduce cross-correlations between all modes. Also note that the applied filter function leads to higher losses in the second mode than in the third mode.


\section{Basis optimization after filtering}\label{sec:epr-entanglement_generation}
From the discussion in Sec. \ref{sec:filtering} it seems that the filtering only has a detrimental impact on the PDC state. It introduces additional losses and correlations between the different modes. However, this filtering process enables us to create single-mode quantum states featuring EPR entanglement. In Fig. \ref{fig:filtered_epr_squeezing} and Fig. \ref{fig:pdc_cov_mat_filt} this effect is not visible, due to the fact that we are still regarding the state in the original broadband mode or \textit{Schmidt basis}, while the filtering reshaped and restructured the spectral properties of the PDC state.

Similarly to the \textit{Schmidt basis}, which reveals the EPR squeezing in the original EPR state we now require a new \textit{effective Schmidt basis} to reveal the suppression of higher-order modes and the remaining EPR entanglement for the filtered state, i.e. we have to move into a new reference frame which unveils the EPR-squeezing properties of the remaining photons and minimizes the correlations between the different modes. From a physical point of view this means that we observe or measure the state in a different set of modes, which is adapted to the distortions introduced by the filters.

We developed two different approaches to obtain this \textit{effective Schmidt basis} via a basis optimization. In Sec. \ref{sec:svd_optimization_routine} we present a simple and straightforward method to find the optimal modes labeled \textit{SVD basis optimization}. This approach, however, is based on heuristic arguments. To verify that it indeed yields optimal results we present a rigorous optimization method labeled \textit{global basis optimization} in Sec. \ref{sec:global_basis_optimization}. This approach, is extremely complicated and computationally challenging, but confirms that our straightforward model from Sec. \ref{sec:svd_optimization_routine} indeed delivers optimal results.

\subsection{Singular-value decomposition basis optimization}\label{sec:svd_optimization_routine}

\begin{figure}[htb]
    \begin{center}
        \includegraphics[width=0.9\linewidth]{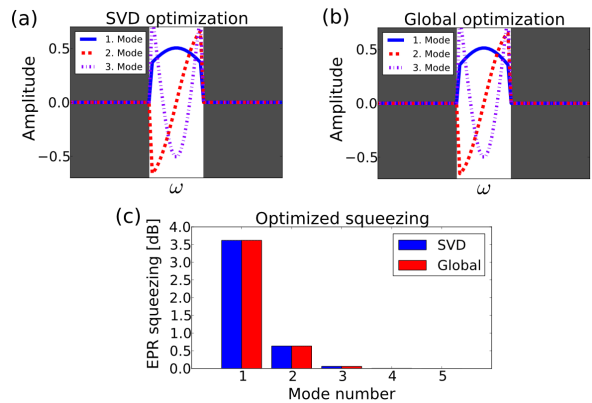}
    \end{center}
    \caption{New basis set (a) obtained via the \textit{SVD basis optimization} and (b) the \textit{global basis optimization} routine. (c) Comparison of the obtained EPR-squeezing values. Both approaches deliver virtually identical results.}
    \label{fig:opt_vs_svd}
\end{figure}

Our goal is to find a simple and straightforward method to obtain a new \textit{effective Schmidt basis} which optimally describe the optical modes of the PDC state after the filtering operations. In Sec. \ref{sec:type-II_PDC} we elaborated that the original \textit{Schmidt basis} is obtained by performing a singular-value decomposition (SVD) of the JSA. Intuitively by simply decomposing the JSA multiplied by the applied filter functions we should obtain mode shapes adapted to the filtering process. To be precise we perform the following SVD (compare Eq. \eref{eq:schmidt_decomposition})
\begin{align}
    T_a(\omega_s) T_b(\omega_i) \left[- \frac{\imath}{\hbar} B f(\omega_s, \omega_i) \right] = \sum_k r_k' \psi'^*_k(\omega_s) \phi'^*_k(\omega_i),
    \label{eq:filter_svd}
\end{align}
and use the obtained mode functions \(\psi'_k(\omega_s)\) and \(\phi'_k(\omega_i)\) to describe the filtered PDC state. Effectively, the only difference to the standard decomposition is the additional multiplication of the JSA by the filter functions, which renders this approach extremely straightforward.

\begin{figure}[htb]
    \begin{center}
        \includegraphics[width=0.9\linewidth]{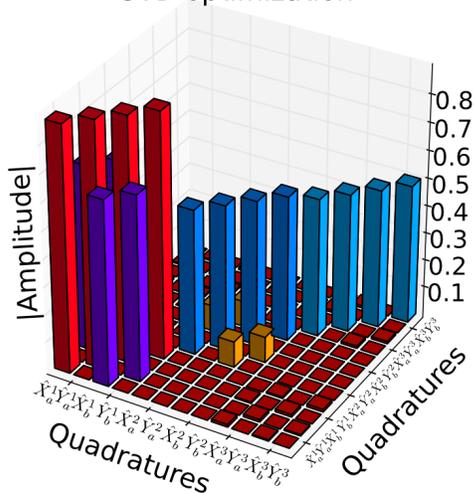}
    \end{center}
    \caption{Covariance matrix of the filtered PDC state from Fig. \ref{fig:pdc_cov_mat_filt} using the \textit{SVD basis optimization}. Performing this basis optimization reveals that filtering effectively suppresses higher-order frequency modes.}
    \label{fig:cov_mat_svd}
\end{figure}

The obtained signal modes, for the considered scenario, are depicted in Fig. \ref{fig:opt_vs_svd}(a). As expected they are much narrower than the original modes and fully located inside the filter bandwidth.

While there is no direct mathematical proof that this approach is optimal, it intuitively makes sense and indeed delivers very good results, as can be seen in Fig. \ref{fig:opt_vs_svd}(c). In comparison to using the original basis to measure EPR squeezing, as depicted in Fig. \ref{fig:filtered_epr_squeezing}, using the optimized basis the main part of the EPR squeezing is contained in the very first optical mode, i.e. filtering indeed effectively suppressed higher-order modes.

The covariance matrix representation of the filtered PDC state in the optimized basis set is given in Fig. \ref{fig:cov_mat_svd}. In comparison to the covariance matrix of the filtered PDC state in the original basis depicted in Fig. \ref{fig:pdc_cov_mat_filt} it shows that using an optimized basis enables us to suppress cross-correlations between different modes and it most importantly reveals that filtering moves a multimode PDC state towards a single-mode operation.

In total this \textit{SVD basis optimization} routine reveals that it is indeed possible to filter a multi-mode PDC state to suppress the all but one optical mode, given that we use the correct basis representation of the state.

\subsection{Global basis optimization}\label{sec:global_basis_optimization}

While the intuitive approach, presented in Sec. \ref{sec:svd_optimization_routine}, delivers very good results it is not clear if it is indeed optimal. To investigate this we developed a second model optimizing the EPR-squeezing values over all possible basis sets.

Our objective is to find a new orthonormal set of modes for the signal and idler beams which maximize the EPR squeezing after filtering. This should reveal its single-mode character and minimize cross-correlations. For the demonstration purposes in this paper, we are able to simplify this procedure by only optimizing a single set of modes $\{\Phi_k\}$ for signal and idler. This is possible due to the fact, that in our exemplary state the signal and idler modes are identical, except for the fact that the odd modes of idler have an additional factor of \(-1\) [see Fig. \ref{fig:pdc_example} (c)]. Using only this one basis set to express and measure the filtered state means that the EPR squeezing in the even modes, where the original signal and idler modes are identical, is located in the \(\hat{X}^{k}_{(-)}\) and \(\hat{Y}^{k}_{(+)}\) quadratures. Correspondingly, due to the additional factor of \(-1\) in the odd idler modes, here, the \(\hat{X}^{k}_{(+)}\) and \(\hat{Y}^{k}_{(-)}\) quadratures show squeezing, when we use the optimized basis set (see App. \ref{app:squeezing_after_filtering}). Elaboration on this effect is given in \cite{patera_quantum_2010}.

Effectively we have to find a new set of real valued functions $\{\Phi_k\}$ which maximize the EPR squeezing after the filtering under the orthogonality constraint
\begin{align}
\int \mathrm d \omega \, \Phi_k(\omega) \Phi_{k'}(\omega) = \delta_{k k'} ,
\end{align}
where the $\Phi_k$ take on the role of $f_k$ and $g_k$ and thus determine the broadband operators Eqs.~\eqref{eq:new_broadband_c} and \eqref{eq:new_broadband_d}. In a discretized formulation we are able to rewrite the basis in terms of a matrix
\begin{align}
A = \left( \Phi_1(\omega), \Phi_2(\omega), \dots \right)
\end{align}
with orthonormal columns
\begin{align}
\sum_l A_{lm} A_{ln} = \delta_{mn} .
\label{constr}
\end{align}
If the $\omega$-sampling in frequency space of the basis functions contains $l$ points and only the first $k$ modes are considered to be relevant, $A$ is a $\mathbb{R}^{l\times k}$ matrix.

To find the optimal basis set \(A\) we use a {\em genetic algorithm} (GA)~\cite{eiben_evo_2003}. We choose this algorithm because we require a {\em global} maximum for the EPR squeezing. 

The objective function $S_{(A,k')}$ maps the basis $A$ onto an EPR squeezing value of a certain mode $k'$. The complete method of how to find the EPR squeezing is lengthy but straightforward and is presented in Appendixes~\ref{app:covariance_matrix} and~\ref{app:squeezing_after_filtering}. Conceptionally, we have to optimize a function
\begin{align}
S_{(A,k')}: \mathbb{R}^{lk} \rightarrow \mathbb{R} ,
\end{align}
i.e., the algorithm looks for a maximum in a $(k\cdot l)$ parameter space. Typical values in this paper are $|k|=5$ modes represented on a $|l|=100$ frequency grid, which are sufficient for our demonstration purposes. It has been thoroughly analyzed in \cite{walther_calculus_2011} that the applied GA is well suited for problems of the current dimension. A convenient way to take into account the constraint Eq.~\eqref{constr} is to decompose $A$ into its QR factorization rather than to use $A$ itself. Because $Q\in\mathbb{R}^{l\times k}$ is orthonormal by construction, we can interpret it as basis set, yet being parametrized by the original components of $A$. This means that the components $(A_{ij})$ are used as {\em genes} in the GA, however, the columns of the $Q$ matrix determine the basis set $\{\Phi_k\}$.

$R\in\mathbb{R}^{k\times k}$ is an upper triangular matrix which makes it possible to successively construct the modes. First we manipulate only the first column in $\{A_{l1}\}$ and optimize the EPR squeezing yielding the mode $\Phi_1(\omega)\equiv (Q_{l1})$ and $R_{11}$. We run the algorithm until the increase in the EPR squeezing of the first mode converges to 0. Then, secondly, we keep $\{A_{l1}\}$ (fixing also  $R_{11}$) and change the entries in the second column $\{A_{l2}\}$. This alters $\Phi_2(\omega)\equiv (Q_{l2})$ together with $R_{12}$ and $R_{22}$ and enables us to optimize the EPR squeezing of the second mode separately. We repeat this procedure with each column until all modes have been obtained. The advantage of the successive building strategy is that merely $l$ parameters are changed in each step instead of $l\cdot k$. This results in the following scheme for the $k'$-th mode:
\begin{align}
\{A_{lk'}\} \longrightarrow
\mathrm{QR} \longrightarrow
\mathrm{mode~} \Phi_{k'}(\omega)\equiv (Q_{lk'}) \nonumber \\
\Phi_{k'}(\omega) \longrightarrow \mathrm{calc~squeezing}
\end{align}
We run the GA with a number of $2^8$ individuals having $|\{A_{lk'}\}|=l$ genes for the $k'$-th mode. We choose two parents from this set randomly and procreate by one-point crossover. A mutation can occur with a probability of 2\%. The cycle is repeated a couple of thousand times until the change in squeezing is less than $10^{-4}$.

\begin{figure}[htb]
    \begin{center}
        \includegraphics[width=\linewidth]{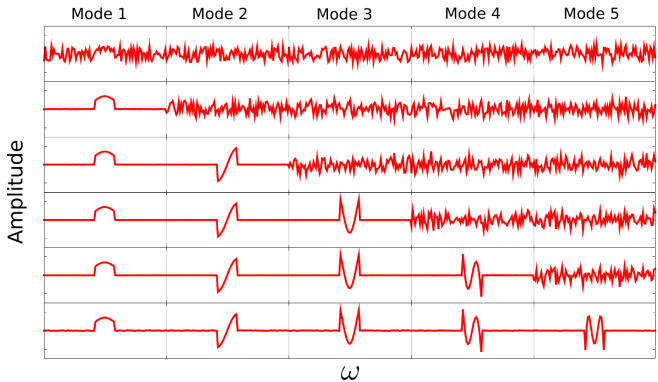}
    \end{center}
    \caption{Evolution of five modes with maximum squeezing obtained by the genetic algorithm. The genes are randomly initialized and the modes 1-5 are found by evolution.}
    \label{fig:evolution}
\end{figure}

The successive procedure is displayed in Fig.~\ref{fig:evolution}. Starting with random data, our algorithm is able to iteratively maximize the EPR squeezing in each mode individually.

It should be noted that the successive maximization of the squeezing of the modes is not the only practicable scheme. In principle, it is desirable to have a large squeezing value in the first mode. Considering only one mode, however, does not provide any information of how much squeezing is left in the higher modes. Maximizing the EPR squeezing in the other modes as well enables us to accurately judge the amount of mode suppression introduced by the filtering.

For the filtered PDC state discussed in Sec. \ref{sec:filtering} we depicted the resulting first three optimized modes in Fig. \ref{fig:opt_vs_svd}(b). The corresponding squeezing values for the individual modes are given by the red bars in Fig. \ref{fig:opt_vs_svd}(c).

The obtained mode shapes from the \textit{Global basis optimization}, depicted in Fig. \ref{fig:opt_vs_svd}(b) are virtually identical to the mode shapes obtained from our \textit{SVD basis optimization} earlier. Similarly the obtained EPR-squeezing distributions in Fig. \ref{fig:opt_vs_svd}(c) do not differ to any noticeable degree. This confirms that our simple and straightforward model from Sec. \ref{sec:svd_optimization_routine} indeed delivers optimal results. The \textit{SVD basis optimization} thus provides us in fact with the \textit{effective Schmidt basis} for the filtered system. Some additional support on why the \textit{SVD basis optimization} approach works so well is given in App. \ref{app:filtered_type-II_PDC}.

\section{Analysis}\label{sec:analysis}

\begin{figure*}[htb]
    \begin{center}
        \includegraphics[width=0.9\textwidth]{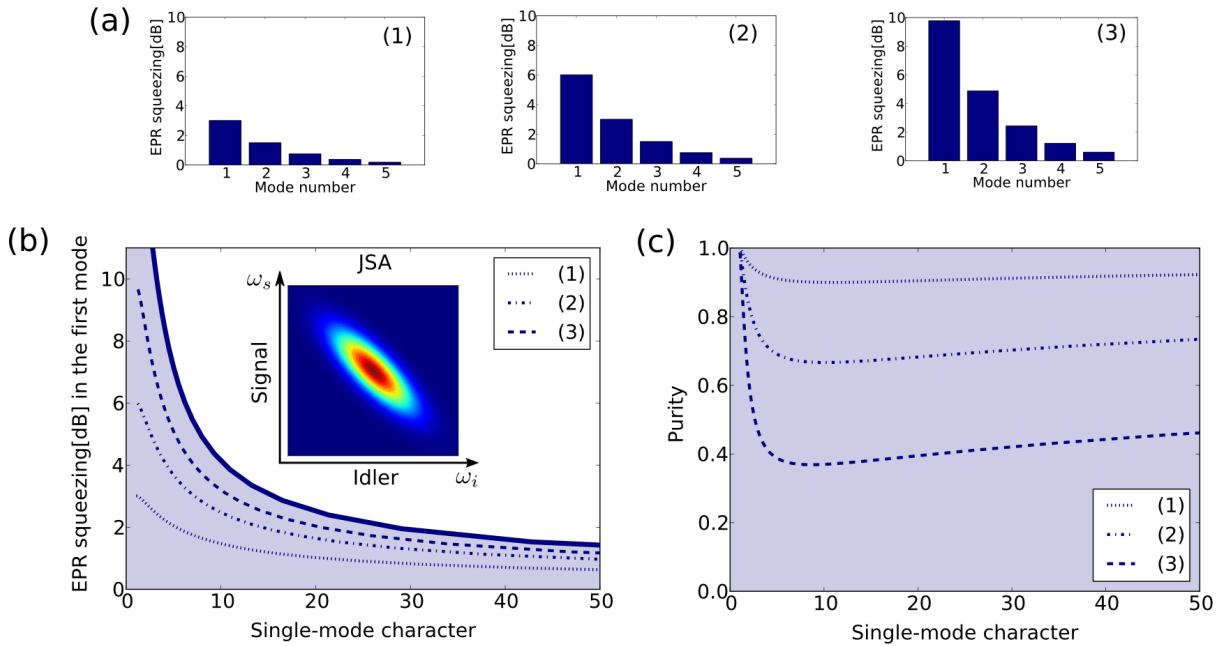}
    \end{center}
    \caption{Quantitative analysis of the impact of filtering on the exemplary PDC state presented in Fig. \ref{fig:pdc_example} using the three different initial EPR-squeezing distributions given in panel (a). (b) Remaining EPR squeezing in the first mode as a function of the single-mode character, for various filter bandwidths. (c) Purity of the filtered state as a function of the single-mode character. The blue shaded areas are the accessible regions using our exemplary PDC state.}
    \label{fig:filtering_analysis}
\end{figure*}

The \textit{SVD basis optimization} developed in Sec. \ref{sec:svd_optimization_routine} finally enables us to accurately quantify the impacts of filtering on PDC in the continuous-variable domain. For this purpose we used our exemplary PDC state from Fig. \ref{fig:pdc_example} and evaluated the remaining EPR squeezing in the first mode, the suppression of the higher-order modes and the purity \cite{adesso_entanglement_2007} of the filtered state for various filter bandwidths and initial EPR-squeezing values. Our results are depicted in Fig. \ref{fig:filtering_analysis}.

For our analysis we used three different initial EPR-squeezing distributions, as depicted in Fig. \ref{fig:filtering_analysis}(a). We visualized the remaining EPR squeezing in the first mode after filtering as a function of the single-mode character of the filtered state in Fig. \ref{fig:filtering_analysis}(b). The three dotted lines correspond to states with the initial EPR-squeezing values presented in Fig. \ref{fig:filtering_analysis}(a). The single-mode character is defined as the ratio between the EPR squeezing in the first mode divided by the EPR squeezing in all higher-order modes. This means a ratio of 10 already corresponds to EPR squeezing in the first mode which is ten times stronger then in all other optical modes. The blue shaded area marks the values accessible using our exemplary PDC spectrum. This figure shows that there is a sharp trade-off between the remaining EPR squeezing and the achievable single-mode character, especially when high initial EPR-squeezing values are present. Interestingly the boundary in Fig. \ref{fig:filtering_analysis}(b) shows that this cannot be offset by higher initial EPR-squeezing values.

The second important parameter of PDC is the remaining purity after the filtering process. We visualized the purity of the filtered states as a function of the single-mode character in Fig. \ref{fig:filtering_analysis}(c). Again we used the exemplary PDC state from Fig. \ref{fig:pdc_example} and the three dotted lines correspond to various filters applied to the initial EPR-squeezing values from \ref{fig:filtering_analysis}(a). It is evident that strongly EPR-squeezed PDC states feature a much higher drop in purity than weakly squeezed states.

In total our analysis shows that filtering PDC enables us to effectively suppress higher-order modes. For highly EPR-squeezed input states the losses in purity and EPR squeezing are, however, severe, whereas weakly EPR-squeezed states only suffer minor losses.

Finally, note that our developed framework is not limited to the symmetric PDC states and identical signal and idler filters, as exemplary presented throughout this paper, but is applicable to all kinds of PDC states and filter configurations.

\section{Conclusion}\label{sec:conclusion}
In conclusion, we developed a simple and straightforward quantitative theoretical model for filtered type-II PDC in the continuous-variable domain.

Our developed \textit{SVD basis optimization} routine provides us with the \textit{effective Schmidt basis} of the filtered state, as verified by our \textit{global basis optimization}. It hence enables the precise and straightforward engineering and evaluation of the resulting filtered PDC states and consequently provides a quantitative analysis tool for the design of experimental implementations. While we found that the global and the SVD optimization yield virtually identical results, it remains an open question to understand if and under what conditions the latter provides the optimal effective Schmidt basis.

Our theoretical framework further enabled us to accurately quantify the impact of narrowband optical filters on type-II PDC. Our analysis shows that narrowband optical filtering of pulsed type-II PDC effectively suppresses all but one optical mode, however, at high EPR-squeezing values the losses in purity and EPR squeezing are severe, whereas the purity and EPR squeezing of weakly squeezed EPR-squeezed states remains mostly unaffected.

This renders filtered PDC optimally suited for experiments in the photon-pair regime, such as the heralding of single-photons, the generation of entangled photon pairs or experiments where low squeezing values are sufficient. However, as soon as high EPR-squeezing values, in a single well-defined optical mode, are required more complicated schemes such as \textit{source engineering} have to be applied.

Finally, it should be noted that our theoretical framework is not only restricted to type-II PDC processes, but can straightforwardly be adapted to type-I PDC and four-wave-mixing processes, due to their similar mathematical structure.

\section{Acknowledgements}
The authors thank Michael Stefszky and Benjamin Brecht for useful discussions and helpful comments.
This work is supported by the Research Training Group 1464 ``Micro- and Nanostructures in Optoelectronics and Photonics''. Cosmo Lupo was supported by the DARPA Quiness Program through US Army Research Office award W31P4Q-12-1-0019.

\appendix

\section{Covariance matrix}\label{app:covariance_matrix}
The covariance matrix of a two-mode quantum state has \(4\times4\) elements for all combinations of the quadratures \(\hat{X}_a, \hat{Y}_a, \hat{X}_b, \hat{Y}_b\), which, for example, can fully describe an EPR state. In our case, we are working with a PDC source, which initially emits \(N\) optical modes, each containing an EPR state, i.e. the covariance matrix is of dimension \(4N \times 4N\). This matrix consists of \(N^2\) \(4\times4\) submatrices, which we label \(a_{kl}\). Each submatrix \(a_{kl}\) describes the correlations between an optical mode \(k\) and another optical mode \(l\). Explicitly written down it is of the form
\begin{widetext}
\begin{align}
    a_{kl} = \frac{1}{2}
    \begin{pmatrix}
        \langle \hat{X}_a^k \hat{X}_a^l + \hat{X}_a^l \hat{X}_a^k \rangle & \langle \hat{X}_a^k \hat{Y}_a^l + \hat{Y}_a^l \hat{X}_a^k \rangle & \langle \hat{X}_a^k \hat{X}_b^l + \hat{X}_b^l \hat{X}_a^k \rangle & \langle \hat{X}_a^k \hat{Y}_b^l + \hat{Y}_b^l \hat{X}_a^k \rangle\\
        \langle \hat{Y}_a^k \hat{X}_a^l + \hat{X}_a^l \hat{Y}_a^k \rangle & \langle \hat{Y}_a^k \hat{Y}_a^l + \hat{Y}_a^l \hat{Y}_a^k \rangle & \langle \hat{Y}_a^k \hat{X}_b^l + \hat{X}_b^l \hat{Y}_a^k \rangle & \langle \hat{Y}_a^k \hat{Y}_b^l + \hat{Y}_b^l \hat{Y}_a^k \rangle\\
        \langle \hat{X}_b^k \hat{X}_a^l + \hat{X}_a^l \hat{X}_b^k \rangle & \langle \hat{X}_b^k \hat{Y}_a^l + \hat{Y}_a^l \hat{X}_b^k \rangle & \langle \hat{X}_b^k \hat{X}_b^l + \hat{X}_b^l \hat{X}_b^k \rangle & \langle \hat{X}_b^k \hat{Y}_b^l + \hat{Y}_b^l \hat{X}_b^k \rangle\\
        \langle \hat{Y}_b^k \hat{X}_a^l + \hat{X}_a^l \hat{Y}_b^k \rangle & \langle \hat{Y}_b^k \hat{Y}_a^l + \hat{Y}_a^l \hat{Y}_b^k \rangle & \langle \hat{Y}_b^k \hat{X}_b^l + \hat{X}_b^l \hat{Y}_b^k \rangle & \langle \hat{Y}_b^k \hat{Y}_b^l + \hat{Y}_b^l \hat{Y}_b^k \rangle
    \end{pmatrix},
    \label{eq:sub_covariance_matrix}
\end{align}
where we dropped the displacements, since all quantum states, considered in this paper, are centered about zero in phase-space. The elements of the covariance matrix of a filtered type-II PDC state are governed by many symmetries. For our filtered PDC state, defined in Eq. \eref{eq:pdc_state_filtered_broadband}, we are able to write the individual submatrices \(a_{kl}\) as
\begin{align}
    a_{kl} = \frac{1}{2}
    \begin{pmatrix}
        a & c & e & g \\
        -c & a & g & -e \\
        f & h & b & d \\
        h & -f & -d & b
    \end{pmatrix}.
    \label{eq:sub_covariance_matrix_2}
\end{align}
with the individual elements defined as
\begin{align}
    \nonumber
    a =& \frac{1}{2}\left(\int \mathrm d \omega \, U_a^k(\omega) U_a^{l*}(\omega) + \int \mathrm d \omega \, R_a^k(\omega) R_a^{l*}(\omega) + \int \mathrm d \omega \, V_a^{k*}(\omega)V_a^l(\omega) \right. \\
    &+ \left. \int \mathrm d \omega \, U_a^l(\omega) U_a^{k*}(\omega) + \int \mathrm d \omega \, R_a^l(\omega) R_a^{k*}(\omega) + \int \mathrm d \omega \, V_a^{l*}(\omega)V_a^k(\omega) \right) \\
    \nonumber
    b =& \frac{1}{2}\left(\int \mathrm d \omega \, U_b^k(\omega) U_b^{l*}(\omega) + \int \mathrm d \omega \, R_b^k(\omega) R_b^{l*}(\omega) + \int \mathrm d \omega \, V_b^{k*}(\omega)V_b^l(\omega) \right. \\
    &+ \left. \int \mathrm d \omega \, U_b^l(\omega) U_b^{k*}(\omega) + \int \mathrm d \omega \, R_b^l(\omega) R_b^{k*}(\omega) + \int \mathrm d \omega \, V_b^{l*}(\omega)V_b^k(\omega) \right) \\
    \nonumber
    c =& \frac{1}{2\imath}\left( - \int \mathrm d \omega \, U_a^k(\omega) U_a^{l*}(\omega) - \int \mathrm d \omega \, R_a^k(\omega) R_a^{l*}(\omega) + \int \mathrm d \omega \, V_a^{k*}(\omega) V_a^l(\omega) \right. \\
    &+ \left. \int \mathrm d \omega \, U_a^l(\omega) U_a^{k*}(\omega) + \int \mathrm d \omega \, R_a^l(\omega) R_a^{k*}(\omega) - \int \mathrm d \omega \, V_a^{l*}(\omega) V_a^k(\omega) \right)\\
    \nonumber
    d =& \frac{1}{2\imath}\left( - \int \mathrm d \omega \, U_b^k(\omega) U_b^{l*}(\omega) - \int \mathrm d \omega \, R_b^k(\omega) R_b^{l*}(\omega) + \int \mathrm d \omega \, V_b^{k*}(\omega) V_b^l(\omega) \right. \\
    &+ \left. \int \mathrm d \omega \, U_b^l(\omega) U_b^{k*}(\omega) + \int \mathrm d \omega \, R_b^l(\omega) R_b^{k*}(\omega) - \int \mathrm d \omega \, V_b^{l*}(\omega) V_b^k(\omega) \right)\\
    e =& \frac{1}{2} \left( \int \mathrm d \omega \, U_a^k(\omega) V_b^l(\omega) + \int \mathrm d \omega \, V_a^{k*}(\omega) U_b^{l*}(\omega) + \int \mathrm d \omega \, U_b^l(\omega) V_a^k(\omega) + \int \mathrm d \omega \, V_b^{l*}(\omega) U_a^{k*}(\omega) \right)\\
    f =& \frac{1}{2} \left( \int \mathrm d \omega \, U_b^k(\omega) V_a^l(\omega) + \int \mathrm d \omega \, V_b^{k*}(\omega) U_a^{l*}(\omega) + \int \mathrm d \omega \, U_a^l(\omega) V_b^k(\omega) + \int \mathrm d \omega \, V_a^{l*}(\omega) U_b^{k*}(\omega) \right)\\
    g =& \frac{1}{2\imath}\left( \int \mathrm d \omega \, U_a^k(\omega) V_b^l(\omega) - \int \mathrm d \omega \, V_a^{k*}(\omega) U_b^{l*}(\omega) + \int \mathrm d \omega \, U_b^l(\omega) V_a^k(\omega) - \int \mathrm d \omega \, V_b^{l*}(\omega) U_a^{k*}(\omega)\right)\\
    h =& \frac{1}{2\imath}\left( \int \mathrm d \omega \, U_b^k(\omega) V_a^l(\omega) - \int \mathrm d \omega \, V_b^{k*}(\omega) U_a^{l*}(\omega) + \int \mathrm d \omega \, U_a^l(\omega) V_b^k(\omega) - \int \mathrm d \omega \, V_a^{l*}(\omega) U_b^{k*}(\omega)\right)
    \label{eq:sub_covariance_matrix_elements}
\end{align}
\end{widetext}

\section{Einstein-Podolsky-Rosen squeezing after filtering}\label{app:squeezing_after_filtering}
Without filtering, the generated EPR squeezing and anti-squeezing can be directly calculated from the \(r_k\)-values. However, when filtering is applied, the formalism becomes more complicated. In this configuration we explicitly have to consider the different variances between the signal and idler beams. They are defined as
\begin{align}
    \nonumber
    &\Delta^{2}\hat{X}_{(-)}^k = \Delta^2\left(\hat{X}_a^k - X_b^k\right) \\
    \nonumber
    &\Delta^{2}\hat{X}_{(+)}^k = \Delta^2\left(\hat{X}_a^k + X_b^k\right) \\
    \nonumber
    &\Delta^{2}\hat{Y}_{(+)}^k = \Delta^2\left(\hat{Y}_a^k + Y_b^k\right) \\
    &\Delta^{2}\hat{Y}_{(+)}^k = \Delta^2\left(\hat{Y}_a^k + Y_b^k\right).
\end{align}
We can directly extract these values from our filtered covariance matrices detailed in App. \ref{app:covariance_matrix}. The variances in mode \(k\) can be calculated from the submatrix \(a_{kk}\), defined in Eq. \eref{eq:sub_covariance_matrix_2}, via the relation
\begin{align}
    \nonumber
    &\Delta^{2}\hat{X}_{(-)}^k = \Delta^{2}\hat{Y}_{(+)}^k = a + b - e - f \\
    &\Delta^{2}\hat{X}_{(+)}^k = \Delta^{2}\hat{Y}_{(-)}^k = a + b + e + f.
\end{align}
These variances can be transformed to the EPR (anti-)squeezing in dB by the formulas
\begin{align}
    \nonumber
    &\mathrm{(anti-)squeezing}[\mathrm{dB}] = -10 \log_{10}\left[\Delta^{2}\hat{X}^{k}_{(+/-)} \right]\\
    &\mathrm{(anti-)squeezing}[\mathrm{dB}] = -10 \log_{10}\left[\Delta^{2}\hat{Y}^{k}_{(+/-)} \right].
    \label{eq:squeezing_db_conversion}
\end{align}

\section{Simulated PDC state}\label{app:simulated_pdc_state}
In the scope of this paper we investigate the effects of filtering on an exemplary anticorrelated type-II PDC state. In order to simplify the discussion we developed a PDC toy model. According to Sec. \ref{sec:type-II_PDC} we only require two sets of mode functions \(\{\psi_k(\omega_s)\}\) and \(\{\phi_k(\omega_i)\}\) for the signal and idler modes respectively and a \(r_k\)-distribution to fully describe a type-II PDC state.

In order to obtain the mode functions, we approximate the JSA \(f(\omega_s, \omega_i)\) as a real two-dimensional normalized Gaussian function --- this corresponds to PDC pumped by a pulsed pump laser --- by using
\begin{align}
    \nonumber
    f(\omega_s, \omega_i) &= \frac{1}{\sqrt{N}}\exp\left[-\frac{\left[\omega_s \cos(\theta) + \omega_i \sin(\theta)\right]^2}{2 \sigma^2_{a}}\right] \\
    &\times \exp\left[-\frac{\left[- \omega_s \sin(\theta) + \omega_i \cos(\theta)\right]^2}{2 \sigma^2_{b}}\right].
    \label{eq:simulated_jsa}
\end{align}
Here \(\sigma_a\) and \(\sigma_b\) give the widths of the individual one-dimensional (1D) Gaussians, and \(\theta\) gives the tilt in the \(\omega_s\)-\(\omega_i\)-plane, and \(\frac{1}{\sqrt{N}}\) is the normalization constant. For our simulations we use \(\sigma_a = 6.0\), \(\sigma_b = 2.0\) and \(\theta= -\frac{\pi}{4}\).

Via a \textit{Schmidt decomposition} we decompose the JSA as
\begin{align}
    f(\omega_s, \omega_i) = \sum_k \lambda_k \psi_k(\omega_s) \phi_k(\omega_i).
    \label{eq:simulated_schmidt_decomposition}
\end{align}
which yields the required signal and idler basis sets and a normalized \(\lambda_k\)-distribution (\(\sum_k \lambda_k^{2}=1\)). Finally we transform the \(\lambda_k\)-distribution to the missing \(r_k\)-distribution via the optical gain \(B\) (\(r_k = B \lambda_k\)), where \(B\) is real valued, positive and adjusted to yield the desired EPR-squeezing values.

This simplified type-II PDC model is extremely flexible, simple and, most importantly, enables us to work with strictly real valued functions, which facilitate a straightforward display of our results throughout the paper.

\section{Properties of filtered type-II parametric down-conversion}\label{app:filtered_type-II_PDC}

In the main part of the paper we state that the filtering, in general, leads to correlations between the individual modes, which, even performing a basis optimization can only be minimized. We also claim that, in accordance with the literature \cite{terhal_schmidt_2000, eisert_schmidt_2001, aniello_class_2008} the resulting mixed quantum states do not, in general, feature a Schmidt decomposition. In this section we are going to perform some analysis concerning these properties of filtered PDC.

In general we would like to find a broadband basis set in which we are able to write the filtered PDC state similar to Eq. \eref{eq:pdc_state_heisenberg_broadband}, i.e. the individual modes are completely independent from each other \cite{braunstein_squeezing_2005}. To illustrate the issues with this transformation let us first revisit Eq. \eref{eq:pdc_state_heisenberg} by using the definitions  for the \(U\) and \(V\) matrices from Eq. \eref{eq:U_V_matrices}
\begin{widetext}
\begin{align}
    \nonumber
    &\hat{a}^{(\mathrm{out})}(\omega) = \int \mathrm d \omega' \sum_k \psi_k^*(\omega) \cosh(r_k) \psi_k(\omega') \hat{a}^{(\mathrm{in})}(\omega') + \int \mathrm d \omega' \sum_k \psi_k^*(\omega) \sinh(r_k) \phi_k^*(\omega') \hat{b}^{(\mathrm{in})\dagger}(\omega')\\
    &\hat{b}^{(\mathrm{out})}(\omega) = \int \mathrm d \omega' \sum_k \phi_k^*(\omega) \cosh(r_k) \phi_k(\omega') \hat{b}^{(\mathrm{in})}(\omega') + \int \mathrm d \omega' \sum_k \phi_k^*(\omega) \sinh(r_k) \psi_k^*(\omega') \hat{a}^{(\mathrm{in})\dagger}(\omega').
    \label{eq:pdc_schmidt_form}
\end{align}
\end{widetext}
To transform this equation into the broadband mode picture we replace the mode functions to the right of the \(\cosh\) and \(\sinh\) terms, in conjunction with the photon creation and destruction operators and the \(\omega'\)-integrals, with the broadband mode operators defined in Eq. \eref{eq:broadband_modes}. To obtain broadband modes on the left hand side of Eq. \eref{eq:pdc_schmidt_form} we multiply, in the case of the upper formula, both sides with \(\int \mathrm d \omega \, \psi_k(\omega)\). This yields broadband modes on the left-hand side and \(\delta_{kl}\) functions on the right hand side getting rid of the summation. We arrive at Eq. \eref{eq:pdc_state_heisenberg_broadband}, which nicely depicts the EPR properties of type-II PDC in the Heisenberg picture.

Unfortunately, in the filtering, case this procedure is not possible any more. If we let the filter from Eq. \eref{eq:filter_operation} act on the PDC state in the Heisenberg picture from Eq. \eref{eq:pdc_state_heisenberg} and use the \textit{Schmidt} form of the \(U\) and \(V\) matrices from Eq. \eref{eq:U_V_matrices}, we arrive at
\begin{widetext}
\begin{align}
    \nonumber
    &\hat{a}^{(\mathrm{out})}(\omega) = T_a(\omega) \left[ \int \mathrm d \omega' \sum_k \psi_k^*(\omega) \cosh(r_k) \psi_k(\omega') \hat{a}^{(\mathrm{in})}(\omega') + \sum_k \psi_k^*(\omega) \sinh(r_k) \phi_k^*(\omega') \hat{b}^{(\mathrm{in})\dagger}(\omega') \right] + R_a(\omega) \hat{v}_a(\omega)\\
    &\hat{b}^{(\mathrm{out})}(\omega) = T_b(\omega) \left[ \int \mathrm d \omega' \sum_k \phi_k^*(\omega) \cosh(r_k) \phi_k(\omega') \hat{b}^{(\mathrm{in})}(\omega') + \sum_k \phi_k^*(\omega) \sinh(r_k) \psi_k^*(\omega') \hat{a}^{(\mathrm{in})\dagger}\omega') \right] + R_b(\omega) \hat{v}_b(\omega).
    \label{eq:filtered_pdc_schmidt_form}
\end{align}
\end{widetext}
This formula clearly shows that the functions \(\psi_k^*(\omega)\) and \(\phi_k^*(\omega)\), which form the detection modes, to the left of the \(\sinh\) and \(\cosh\) terms are multiplied by the filter shapes. If we repeat the steps as for the unfiltered PDC state and multiply, in the upper formula, both sides with \(\int \mathrm d \omega \, \psi_k(\omega)\) the added filter functions lead to overlaps between several modes simultaneously. Consequently couplings between the different modes occur. It is not possible to find a new broadband mode basis, where the different modes are uncorrelated.

There is only a single exception: If we assume that the signal and idler modes are real and identical, all \(r_k\)-values share an identical excitation and the filters in the signal and idler are are identical as well. In this specific case we can write the filtered state as
\begin{widetext}
\begin{align}
    \nonumber
    &\hat{a}^{(\mathrm{out})}(\omega) = \int \mathrm d \omega'  \cosh(r) \sum_k T(\omega) \psi_k(\omega) \psi_k(\omega') \hat{a}^{(\mathrm{in})}(\omega') + \sinh(r) \sum_k T(\omega) \psi_k(\omega) \psi_k(\omega') \hat{b}^{(\mathrm{in})\dagger}(\omega') + R(\omega) \hat{v}_a(\omega)\\
&\hat{b}^{(\mathrm{out})}(\omega) = \int \mathrm d \omega'  \cosh(r) \sum_k  T(\omega) \psi_k(\omega) \psi_k(\omega') \hat{b}^{(\mathrm{in})}(\omega') + \sinh(r) \sum_k  T(\omega) \psi_k(\omega) \psi_k(\omega') \hat{a}^{(\mathrm{in})\dagger}(\omega') + R(\omega) \hat{v}_b(\omega).
    \label{eq:filtered_pdc_schmidt_form_2}
\end{align}
\end{widetext}
In this simplification the same term appears four times, on which we now perform the following Schmidt decomposition
\begin{align}
    \sum_k  T(\omega) \psi_k(\omega) \psi_k(\omega') = \sum_k \kappa_k \varphi_k(\omega) \xi_k(\omega').
    \label{eq:schmidt_decomposition_filter}
\end{align}
Further introducing the four new broadband mode functions
\begin{align}
    \nonumber
    \hat{E}^{(\mathrm{out})}_k = \int \mathrm d \omega \, \varphi_k(\omega) \hat{a}(\omega)\\
    \nonumber
    \hat{F}^{(\mathrm{out})}_k = \int \mathrm d \omega \, \varphi_k(\omega) \hat{b}(\omega)\\
    \nonumber
    \hat{G}^{(\mathrm{in})}_k = \int \mathrm d \omega \, \xi_k(\omega) \hat{a}(\omega)\\
    \hat{H}^{(\mathrm{in})}_k = \int \mathrm d \omega \, \xi_k(\omega) \hat{b}(\omega),
    \label{eq:broadband_modes filter}
\end{align}
we are, in fact, able to write Eq. \eref{eq:filtered_pdc_schmidt_form_2} into the broadband mode formalism and decouple the individual modes
\begin{align}
    \nonumber
    \hat{E}^{(\mathrm{out})}_k = \kappa_k\left[\cosh(r) \hat{G}^{(\mathrm{in})}_k + \sinh(r) \hat{H}^{(\mathrm{in})\dagger}_k\right] + R(\omega) \hat{v}_a(\omega)\\
    \hat{F}^{(\mathrm{out})}_k = \kappa_k\left[\cosh(r) \hat{H}^{(\mathrm{in})}_k + \sinh(r) \hat{G}^{(\mathrm{in})\dagger}_k\right] + R(\omega) \hat{v}_a(\omega).
    \label{eq:filtered_pdc_broadband}
\end{align}
In this approximation it is consequently possible to find a new broadband-mode basis, where all individual modes remain orthonormal after the filtering and the filter, in fact, simply acts as a standard loss, which, however, affects the individual modes differently.

For actual PDC sources it is, however, not possible to actually implement all of the above mentioned simplifications, with only the exception of a PDC process pumped by a continuous-wave laser which can approximate these requirements. Hence our conclusion from the main part of the paper remains valid: Filtering, in general, leads to correlations between different modes. Still this analytic calculation gives some insight into our \textit{SVD basis optimization} routine, presented in the main part of the paper. It is, in fact, almost identical to the decomposition shown in Eq. \eref{eq:schmidt_decomposition_filter}, which partly explains why its performance is almost indistinguishable from the \textit{global basis optimization} routine.

\bibliography{filtering_PDC.bib}

\end{document}